\documentclass[12pt]{article}
\pdfoutput=1
\usepackage[utf8]{inputenc}
\usepackage[T1]{fontenc}
\usepackage{graphicx}
\usepackage{amsmath}
\usepackage{amssymb}
\usepackage{tensor} 
\usepackage{braket}
\usepackage{comment}
\usepackage{mathtools}
\usepackage{caption}
\usepackage{cite}

\def\aff#1{\vspace{-12pt}{\normalsize #1}}

                \def\inn{\hook}
               
          \def\ads3{{\rm AdS$_3$}}

\def\bull{\raise.25ex\hbox{\vrule height.8ex width.8ex}}

\def\Lie{{\cal L}\hspace{-.7em}\raise.25ex\hbox{--}\hspace{.2em}}

\def\mb#1{\hbox{{\boldmath $#1$}}}

\def\hook{\hbox{\vrule height0pt width4pt depth0.3pt
\vrule height7pt width0.3pt depth0.3pt
\vrule height0pt width2pt depth0pt}\hspace{0.8pt}}

          \def\cE{{\cal E}}

                     \def\cE{{\cal E}}

     \def\bcE=\mb{\cE}

\def\be{\begin{equation}}             \def\ee{\end{equation}}
\def\ba#1{\begin{array}{#1}}          \def\ea{\end{array}}
\def\bea{\begin{eqnarray} }           \def\eea{\end{eqnarray} }
\def\beann{\begin{eqnarray*} }        \def\eeann{\end{eqnarray*} }
\def\beal{\begin{eqalign}}            \def\eeal{\end{eqalign}}
             
\def\bsubeq{\begin{subequations}}     \def\esubeq{\end{subequations}}
\def\bitem{\begin{itemize}}           \def\eitem{\end{itemize}}

\def\aff#1{{\normalsize #1}}

\title{Near-horizon geometry with torsion: Kerr-AdS black hole}
\author{B. Cvetkovi\'c  and D. Rakonjac\footnote{
        Email addresses: \texttt{cbranislav@ipb.ac.rs, danilor@ipb.ac.rs}} \\
\aff{Institute of Physics, University of Belgrade,
                           Pregrevica 118, 11080 Belgrade, Serbia} }
\date{}

\begin{document}

\maketitle

\begin{abstract}
We consider the general construction of near-horizon limit for extremal black hole solutions with non-trivial torsion, and derive the covariant geometric conditions for the existence this limit. A near-horizon solution with torsion is constructed in the case of extremal Kerr-AdS black hole, starting from an ansatz compatible with near-horizon symmetry. We demonstrate that this solution corresponds to the limit of a new black hole solution with torsion.
\end{abstract}

\section{Introduction}

The calculations of black hole entropy continue to be a fruitful area of research, linking established results in classical black hole physics with ongoing studies in quantum gravity. In recent years, a Hamiltonian method for calculating black hole entropy within the first-order formalism has been proposed \cite{BlagojevicCvetkovicEntropyHamiltonApproach}. This approach is based on the concept developed by Wald \cite{WaldEntropy}, which defines black hole entropy as a conserved charge on the horizon. Developed within the first-order formalism, this method has broad applications, especially in the context of Riemann-Cartan geometry. Its validity has been confirmed through calculations of black hole entropies across various solutions in Poincar\'e gauge theory (PG) and general relativity (GR)\cite{BlagojevicCvetkovicPGKerr,BlagojevicCvetkovicKerrAdS,BlagojevicCvetkovicRNlike,BlagojevicCvetkovicKNAdS}. However, this approach is limited to non-extremal black holes. It relies on the variational principle, deriving entropy through the first law of black hole mechanics, $T\delta S = \delta\Gamma_H$, where $\Gamma_H$ is the boundary term on the black hole horizon. For
extremal black holes(where $T=0$), this relation is identically satisfied, and thus, it is not possible to derive entropy from it.

In the study of extremal black holes, an alternative approach was developed \cite{GuicaKerrCFT,CompereKerrCFT,CarlipBHEntropyCFT}. This method relies on the fact that extremal black holes possess a well-defined near-horizon geometry \cite{BardeenHorowitzNHEK,KunduriLuciettiNHSym,KunduriNH}, which exhibits enhanced symmetry compared to the original black hole solution. In particular, the near-horizon geometry possesses a conformal symmetry, leading to an effective conformal field theory description of black hole entropy \cite{CarlipBHEntropyCFT,KunduriLuciettiNHSym}. The canonical approach to calculating black hole entropy within the first-order formalism has been applied to extremal black holes in the context of PG for several simple models \cite{CvetkovicRakonjacNHEK,CvetkovicRakonjacSpacelikeStretched}, by performing a canonical analysis of their near-horizon geometries. This method proved successful, motivating further investigation into the entropy of extremal black holes with nontrivial torsion. However, it was found that near-horizon geometries—guaranteed to exist for extremal black holes in general relativity—are often ill-defined for most solutions of interest in PG. The issue arises because, while the near-horizon metric remains well defined as expected in the extremal case, the torsion tensor typically diverges. In this paper, we initiate a study of this problem by deriving the conditions necessary for the existence of near-horizon geometries with nontrivial torsion and by presenting a new black hole solution that avoids the pathology of an ill-defined near-horizon limit in the extremal case.

The paper is organized as follows. Section 2 provides a review of constructing a near-horizon geometry for extremal black holes in the general case. This discussion is then extended to include solutions with arbitrary torsion tensors. Section 3 presents the Kerr-AdS solution with torsion as a specific example. It is observed that this solution does not possess a well-defined near-horizon limit. In Section 4, using the well-known symmetries of near-horizon black holes, we construct an ansatz that extends these symmetries to include the torsion tensor. The aim is to find a near-horizon geometry corresponding to the Kerr-AdS metric. Section 5 utilizes the double duality method to solve the field equations, yielding a near-horizon geometry of the Kerr-AdS metric with non-trivial torsion. A key question that arises is how to find a Kerr-AdS black hole solution with torsion that corresponds to this near-horizon limit in the extremal case. This solution is presented in Section 6.

The conventions in the paper are the same as in the ref. \cite{BlagojevicCvetkovicEntropyHamiltonApproach}. Three types of indices are used over the course of the paper. The Latin indices $(i,j,\dots)$ are the local Lorentz indices, the first Greek indices $(\alpha, \beta, \dots)$ are corresponding to coordinate components on a hypersurface, while the second part of Greek indices $(\mu,\nu,\dots)$ correspond to coordinate components on the spacetime manifold. The signature of the metric is $\eta_{ij} = (1,-1,-1,-1)$, and $\varepsilon_{ijkl}$ is the totally antisymmetric Levi-Civita symbol with $\varepsilon_{0123}=1$.


\section{General construction of a near horizon limit}
\setcounter{equation}{0}

\subsection{Extremal black holes in general relativity}
We here repeat the construction made in \cite{KunduriNH,FriedrichRaczWald,MoncriefIsenberg}, for the general form of near horizon limit of a degenerate Killing horizon. It will be seen that extremality is the neccessary and sufficient
condition for the existence of near-horizon geometry when torsion tensor is equal to zero.
\par Consider a null hypersurface $\mathcal{H}$ in spacetime manifold $(M,g)$, such that its null normal vector field is a Killing vector field $k$ of spacetime. Being the null normal, the vector field $k$ is also tangent to the hypersurface,
and it's integral curves are geodesics which generate $\mathcal{H}$. Null hypersurfaces which are generated by Killing vector fields are called Killing horizons, and so the following construction applies to general Killing horizons. By Hawking's
rigidity theorem \cite{HawkingBHinGR,HawkingEllis,ChruscielRigidity}, event horizons of black holes are Killing horizons, so application of the following to event horizons is immediate. In general, these geodesics need not be affinely
parametrized, they satisfy the general form of the geodesic equation $k^\mu\nabla_\mu k^\nu = \kappa k^\nu$, where $\kappa$ is the parameter called surface gravity of the black hole. The case when $\kappa=0$ is called a degenerate Killing horizon,
or in the case of event horizon, such a black hole is called extremal black hole. We now proceed to construct coordinates adapted to the Killing vector field $k$ in the neighborhood of the horizon.
\par If $u$ is parameter along the generators, we can align the coordinates on $\mathcal{H}$ so that $k=\partial_u$. Since integral curves of $k$ generate $\mathcal{H}$ we can consider foliation of spacelike cross-sections $H$ such that $u$ is constant
over $H$ and every generator crosses $H$ exactly once. We choose arbitrary coordinates $x^\alpha$ on these cross-sections in a smooth way, such that $x^\alpha$ are constant along the generators of $\mathcal{H}$. We have thus completed the construction of
coordinates on $\mathcal{H}$. To extend the coordinates to a neighborhood of $\mathcal{H}$ in $M$, we consider at every point in $\mathcal{H}$ a null vector $\ell$ such that $k \cdot \ell =1$ and $\ell \cdot \partial_\alpha = 0$. Consider a congruence of null
geodesics in $M$ crossing $\mathcal{H}$, such that each geodesic crosses $\mathcal{H}$ exactly once, and the vectors $\ell$ are tangent to geodesics on $\mathcal{H}$. Let $r$ be the affine parameter of these geodesics, and let us extend coordinates to a neighbourhood
of $\mathcal{H}$, such that $(u,x^\alpha)$ are constant along the geodesics of this congruence, such that $r=0$ on $\mathcal{H}$. Thus, we have completed the construction of coordinates in a neighborhood of $\mathcal{H}$ in $M$. This construction applies to any
null hypersurface, and these coordinates are usually called Gaussian null coordinates \cite{MoncriefIsenberg}.\\
\par Further, we extend the definition of $k$ and $\ell$ to the neighborhood of $\mathcal{H}$ in $M$ such that $k = \partial_u$ and $\ell = \partial_r$. It holds by construction that:
\begin{equation}
    \ell^\mu\nabla_\mu\ell = 0 \qquad [k,\ell] = 0
\end{equation}
From this, it can easily be proven that $\ell^\mu\nabla_\mu(k\cdot\ell) =0$ and $\ell^\mu\nabla_\mu(\ell\cdot\partial_\alpha) = 0$. Thus, we have that $k \cdot \ell = 1$ and $\ell \cdot \partial_\alpha = 0$ in the whole neighborhood of $\mathcal{H}$.\\
With this in mind, the general metric in this chart in the neighborhood of $\mathcal{H}$ takes form:
\begin{equation}
    ds^2 = rf(r,x^\alpha)du^2 + 2dudr + r h_\alpha(r,x^\beta)dudx^\alpha + \gamma_{\alpha\beta}(r,x^\gamma)dx^\alpha dx^\beta
\end{equation}
where we assume that the functions $f, h_\alpha, \gamma_{\alpha\beta}$ are analytic functions of coordinates.\\
A near-horizon transformation is a scaling transformation of the form:
\begin{equation}
    u \rightarrow \frac{u}{\varepsilon} \qquad r \rightarrow \varepsilon r
\end{equation}
Taking the limit as $\varepsilon \rightarrow 0$, we obtain the near-horizon limit geometry(if it exists). In order for the metric to have a smooth limit, we notice that the following has to hold:
\begin{equation*}
    \partial_r g_{uu}|_{r=0} = 0
\end{equation*}
This condition can be transformed into a covariant condition:
\begin{equation*}
    \partial_r g_{uu} = \ell^\mu \nabla_\mu (k \cdot k)  = 2\ell^\mu(\nabla_\mu k^\nu)k_\nu = 2k^\mu(\nabla_\mu \ell^\nu)k_\nu = 2k^\mu (\nabla_\mu(k\cdot \ell) - (\nabla_\mu k^\nu)\ell_\nu)
\end{equation*}
where we have used (1) and the Leibniz rule to transform the expression.\\
We then evaluate this expression at $r=0$ and use the geodesic equation for Killing vector $k$ to get:
\begin{equation}
    \partial_r g_{uu}|_{r=0} = -2\kappa
\end{equation}
Therefore, the condition for the existence of a smooth near-horizon limit of the metric is the condition for degeneracy of the Killing horizon $\kappa = 0$. In case where Killing horizon is the event horizon
of a black hole, this condition is the condition for Hawking temperature to vanish, that is, for the black hole to be extremal.

\subsection{Near-horizon limit in the presence of torsion}

When torsion is non-zero, we consider our spacetime as a manifold with a general metric-compatible affine connection $(M,g,\Gamma)$. Every metric-compatible connection admits a unique decomposition:
\begin{equation}
    \Gamma^\mu_{\nu\rho} = \tilde{\Gamma}^\mu_{\nu\rho} + K^\mu_{\hphantom{\mu}\nu\rho}
\end{equation}
where $\tilde{\Gamma}^\mu_{\nu\rho}$ is the Levi-Civita connection and $K^\mu_{\hphantom{\mu}\nu\rho}=\frac{1}{2}(T^\mu_{\hphantom{\mu}\nu\rho} - T^{\hphantom{\rho}\mu}_{\rho\hphantom{\mu}\nu} + T^{\hphantom{\mu}\hphantom{\rho}\mu}_{\nu\rho})$ is contorsion tensor, $T^\mu_{\hphantom{\mu}\nu\rho} = 2\Gamma^\mu_{[\nu\rho]}$ being the torsion tensor.\\
It is now easily seen that the spacetime geometry in this case can be uniquely determined by prescribing metric and torsion, instead of metric and connection. If we now consider a Killing horizon in $M$, like we did in the last subsection, the results that
we have obtained remain valid, so we now assume that the surface gravity of the horizon is zero, i.e. the horizon is degenerate. We need a condition that torsion has to satisfy in order to also have a smooth near-horizon limit. We will require that torsion
is analytic and regular in Gaussian null coordinates near the horizon. We then perform the near-horizon transformation and obtain the following behavior of independent torsion components:
\begin{align*}
    &T^r_{\hphantom{r}u\alpha} \rightarrow \frac{1}{\varepsilon^2}T^r_{\hphantom{r}u\alpha} \\
    &T^r_{\hphantom{r}ur} \rightarrow \frac{1}{\varepsilon}T^r_{\hphantom{r}ur} &&T^r_{\hphantom{r}\alpha\beta}\rightarrow \frac{1}{\varepsilon}T^r_{\hphantom{r}\alpha\beta} &&T^\alpha_{\hphantom{\alpha}u\beta}\rightarrow \frac{1}{\varepsilon}T^\alpha_{\hphantom{\alpha}u\beta} \\
    &T^r_{\hphantom{r}r\alpha} \rightarrow T^r_{\hphantom{r}r\alpha} &&T^u_{\hphantom{u}u\alpha}\rightarrow T^u_{\hphantom{u}u\alpha} &&T^{\alpha}_{\hphantom{\alpha}ur} \rightarrow T^{\alpha}_{\hphantom{\alpha}ur} &&T^\alpha_{\hphantom{\alpha}\beta\gamma} \rightarrow T^\alpha_{\hphantom{\alpha}\beta\gamma}\\
    &T^u_{\hphantom{u}ur} \rightarrow \varepsilon T^u_{\hphantom{u}ur} &&T^u_{\hphantom{u}\alpha\beta}\rightarrow \varepsilon T^u_{\hphantom{u}\alpha\beta} &&T^\alpha_{\hphantom{\alpha}r\beta} \rightarrow \varepsilon T^\alpha_{\hphantom{\alpha}r\beta} \\
    &T^u_{\hphantom{u}r\alpha} \rightarrow \varepsilon^2 T^u_{\hphantom{u}r\alpha}
\end{align*}
The components separate in powers of $\varepsilon$ parameter, so we can easily identify the following conditions that assure the smoothness of near-horizon limit.
\begin{equation}
    T^r_{\hphantom{r}u\alpha}|_{r=0} = 0 \quad \partial_rT^r_{\hphantom{r}u\alpha}|_{r=0} = 0 \quad T^r_{\hphantom{r}ur}|_{r=0}=0 \quad T^r_{\hphantom{r}\alpha\beta}|_{r=0} = 0 \quad T^\alpha_{\hphantom{\alpha}u\beta}|_{r=0} = 0
\end{equation}
We want to generalize these conditions to covariant ones.\\
We notice from the construction of Gaussian null coordinates, that if we restrict our coordinate chart to $(u,0,x^\alpha)$, we will obtain a chart intrinsic to the horizon $\mathcal{H}$. Taking now $\{y^\mu\}$ to be an arbitrary chart in the neighborhood
of $\mathcal{H}$ in $M$, we can construct a basis of tangent vectors to $\mathcal{H}$ as:
\begin{equation}
    k^\mu = \left(\frac{\partial y^\mu}{\partial u}\right)_{x^\alpha} \qquad e^\mu_{\hphantom{\mu}\alpha} = \left(\frac{\partial y^\mu}{\partial x^\alpha}\right)_{u}
\end{equation}
The auxilliary null vector field $\ell$ completes this basis at every point to a basis of tangent space of $M$. In the construction of Gaussian null coordinates around the horizon, we note that the basis of the tangent space of $\mathcal{H}$ takes form:
\begin{equation*}
    k^\mu = \delta^\mu_u \qquad e^\mu_{\hphantom{\mu}\alpha}=\delta^\mu_\alpha
\end{equation*}
We can then apply the previously used technique of replacing particular components with contractions with the basis. We obtain the following covariant conditions:
\begin{equation}
\begin{split}
    &T_{\mu\nu\rho}k^\mu k^\nu = 0 \\
    &k_\mu T^\mu_{\hphantom{\mu}\alpha\beta} \equiv k_\mu T^\mu_{\hphantom{\mu}\nu\rho}e^\nu_{\hphantom{\nu}\alpha} e^\rho_{\hphantom{\rho}\beta} = 0 \\
    &T_{\alpha\mu \beta}k^\mu \equiv T_{\nu\mu\rho}k^\mu e^\nu_{\hphantom{\nu}\alpha}e^\nu_{\hphantom{\rho}\beta} = 0\\
    &\ell^\mu\nabla_\mu(T_{\nu\rho\sigma}k^\nu k^\rho e^\sigma_{\hphantom{\sigma}\alpha}) = 0
\end{split}
\end{equation}
where all the equations are evaluated on $\mathcal{H}$. \\
All of the conditions are scalar with respect to changes of coordinates in $M$, and covariant with respect to the change of intrinsic coordinates on $\mathcal{H}$, thus we have obtained the
covariant condition for the existence of near-horizon limit in presence of torsion. In fact, the vector $k$ in the conditions above need not be a Killing vector(it can be any null vector field tangent to the generators), but for our purposes
we will define it to be the Killing vector field that generates the horizon.\\
If we label the basis of the tangent space at a point in $\mathcal{H}$, by $E_{A} \equiv (k,e_\alpha)$, we can define induced torsion tensor in analogy to induced metric as a pullback of torsion to the horizon:
\begin{equation*}
    T_{ABC} = T_{\mu\nu\rho}E^\mu_{\hphantom{\mu}A}E^\nu_{\hphantom{\nu}B}E^\rho_{\hphantom{\rho}C}
\end{equation*}
Then our conditions can be read as restrictions on induced torsion. In particular, the first three conditions restrict the intrinsic torsion to only have non-zero components in directions tangent to the cross section $H$ of $\mathcal{H}$. The first condition
appeared before in \cite{ZerothLawTorsion}, where it was used to extend the proof of zeroth law of black hole thermodynamics to the case of non-zero torsion. It implies that torsion current is zero along the generators of $\mathcal{H}$. In a way, we require with
these conditions that the induced torsion on $\mathcal{H}$ is similarly degenerate to induced metric, having non-zero components only on the cross-sections $H$. The last condition restricts
the transverse derivative of induced torsion. It is unfortunate, however, that most interesting solutions in Poincare gauge theory, describing black holes with dynamical torsion, do not satisfy these conditions. We will discuss this in next sections, and try to offer
a tentative solution to the problem.

\section{An example: Kerr-AdS black hole with torsion}
\setcounter{equation}{0}

When considering spacetimes with torsion, the most natural framework to work in is Poincare gauge theory(PG). For detailed review of the theory, see \cite{BlagojevicBook,BlagojevicHehlPGReview}. The geometric structure of PG is defined by two gauge potentials of the Poincare group,
namely the tetrad and spin connection 1-forms $(\vartheta^i, \omega^{ij})$ and their corresponding field strengths, the torsion and curvature 2-forms $T^i$ and $R^{ij}$ which are given by Cartan structure equations:
\begin{equation}
    T^i = d\vartheta^i + \omega^i_{\hphantom{i}j}\wedge \vartheta^j \qquad R^{ij} = d\omega^{ij} + \omega^i_{\hphantom{i}k}\wedge \omega^{kj}
\end{equation}
The most general gravitational Lagrangian which is parity invariant and at most quadratic in field strengths is given by:
\begin{equation}
    L_G = -^*(a_0R +2\Lambda) + T^i \wedge \sum_{n=1}^3 \prescript{*}{}{(a_n \prescript{(n)}{}{T_i})} + \frac{1}{2}R^{ij}\wedge\sum_{n=1}^{6}\prescript{*}{}{(b_n \prescript{(n)}{}{R_{ij}})}
\end{equation}
where $(a_0, \Lambda, a_n, b_n)$ are the coupling constants, the first term coming from Einstein-Cartan gravity with cosmological constant, and the rest of the terms being constructed out of irreducible parts of torsion and curvature, respectively. The explicit forms
of irreducible components will be given in Appendix A. The theory with this Lagrangian is usually called quadradic PG gravity.\\
Varying the Lagrangian gives the vaccuum field equations which can be written in covariant form as:
\bsubeq
\begin{align}
   \delta \vartheta^i:&  \qquad DH_i + E_i = 0\\
   \delta \omega^{ij}:&  \qquad DH_{ij} + E_{ij} = 0
\end{align}
\esubeq
where $H_i := \partial L_G / \partial T^i$ and $H_{ij} = \partial L_G / \partial R^{ij}$ are the graviatation covariant momenta, and $E_i := \partial L_G / \partial \vartheta^i$ and $E_{ij} = \partial L_G / \partial \omega^{ij}$ are the associated gravitational currents,
and $D$ represents the exterior covariant derivative with connection $\omega^{ij}$.\\

We're going to consider here a solution to field equations (3.3), obtained by Baekler et al. \cite{McCreaBaeklerGuersesPGKerr,BaeklerGuersesHehlMcCreaKerrNewmann1988}, which describes a Kerr-AdS black hole with dynamical torsion. The canonical analysis of the solution was performed in \cite{BlagojevicCvetkovicKerrAdS}, where entropy was
obtained in the non-extremal case. We will see that the near-horizon limit of torsion in this solution is not well-defined.\\
The metric of Kerr-AdS black hole in Boyer-Lindquist coordinates \cite{CarterBHLesHouches} is given by:

\begin{equation}
    ds^2 = N^2\left(dt -\frac{a}{\alpha}\sin^2\theta d\varphi\right)^2 - \frac{dr^2}{N^2} - \frac{d\theta^2}{F^2} - F^2{\rho^2}\left(adt - \frac{(r^2+a^2)}{\alpha}d\varphi\right)^2
\end{equation}
where $N=\sqrt{\frac{\Delta}{\rho^2}}$, $F=\sqrt{\frac{f}{\rho^2}}$ and:
\begin{equation}
\begin{split}
    \Delta(r) &= (r^2 + a^2)(1+\lambda r^2) - 2mr \qquad \alpha = 1 - \lambda a^2\\
    \rho^2(r,\theta) &= r^2 + a^2 \cos^2\theta \qquad f(\theta) = 1 - \lambda a^2 \cos^2\theta
\end{split}
\end{equation}
The parameters, $m$ and $a$ are the parameters of the solution corresponding to the mass and specific angular momentum of the black hole, and $\lambda = -\Lambda/3 :=1/\ell^2$ relates to the AdS radius of the background geometry. The function $\Delta(r)$ determines the
radial coordinate of the horizon, and generically $\Delta(r) = 0$ has two solutions, being inner and outer horizon radii $r_-$ and $r_+$. The extremal case can then be defined as the case when $r_- = r_+$. We obtain the parameters of mass and specific angular momentum in this case
by solving the system of equations $\Delta(r) = 0$, $\partial_r\Delta(r) =0$, and we obtain:
\begin{equation}
	a_{\text{ext}} = r_+\sqrt{\frac{1 + 3\lambda r_+^2}{1 - \lambda r_+^2}} \qquad m_{\text{ext}} = r_+\frac{(1+\lambda r_+^2)^2}{1-\lambda r_+^2}
\end{equation}
The surface gravity of the solution is calculated to be $\kappa = \frac{\partial_r\Delta(r)}{r^2+a^2}$ which is equal to zero in the extremal case as expected.\\
Since metric is put in a diagonal form, we obtain the most natural choice of the orthonormal tetrad:
\begin{equation}
\begin{split}
    \vartheta^0 &= N\left(dt -\frac{a}{\alpha}\sin^2\theta d\varphi\right) \qquad \vartheta^1 = \frac{dr}{N}\\
    \vartheta^2 &= \frac{d\theta}{F} \qquad \vartheta^3 = F\sin\theta\left(adt - \frac{(r^2+a^2)}{\alpha}d\varphi\right)
\end{split}
\end{equation}
The torsion in this frame is given by:
\begin{equation}
\begin{split}
    T^0 &= T^1 = \sqrt{\frac{\rho^2}{\Delta}}(-V_1\vartheta^0 \wedge \vartheta^1 - 2V_4 \vartheta^2\wedge \vartheta^3) + \frac{\rho^2}{\Delta}(V_2(\vartheta^0-\vartheta^1)\wedge \vartheta^2 + V_3(\vartheta^0 - \vartheta^1)\wedge \vartheta^3)\\
    T^2 &= \sqrt{\frac{\rho^2}{\Delta}}(V_5(\vartheta^0 - \vartheta^1)\wedge \vartheta^2 + V_4 (\vartheta^0 - \vartheta^1)\wedge \vartheta^3)\\
    T^3 &= \sqrt{\frac{\rho^2}{\Delta}}(-V_4(\vartheta^0-\vartheta^1)\wedge \vartheta^2 + V_5(\vartheta^0 - \vartheta^1)\wedge \vartheta^3)
\end{split}
\end{equation}
where:
\begin{align*}
    &V_1 = \frac{m}{\rho^4}(r^2 - a^2\cos^2\theta) \qquad V_2 = -\frac{mF}{\rho^4}ra^2\sin\theta\cos\theta\\
    &V_3 = \frac{mF}{\rho^4}r^2a\sin\theta \qquad V_4 = \frac{m}{\rho^4}ra\cos\theta \qquad V_5 = \frac{m}{\rho^4}r^2
\end{align*}

\subsection{Near-horizon limit}
Instead of passing to Gaussian-null coordinates, we write here the near-horizon transformation directly. We expand the metric around the horizon by substituting $r = r_+(1+\varepsilon r')$, and then, using the extremal parameters (3.6), we obtain that:
\begin{equation}
    \Delta(r) \sim \varepsilon^2r_+^2 r'^2 V + \mathcal{O}(\varepsilon^3) \qquad V = \frac{1 + 6\lambda r_+^2 - 3\lambda^2 r_+^4}{1 - \lambda r_+^2}
\end{equation}
The near-horizon limit is given by the transformation:
\begin{equation}
	t = \frac{(r_+^2+a^2)}{V}\frac{t'}{\varepsilon r_+} \qquad r = r_+(1+\varepsilon r') \qquad \varphi = \varphi' + \frac{a\alpha}{V}\frac{t'}{\varepsilon r_+}
\end{equation}
along with the limit $\varepsilon \rightarrow 0$ taken after the transformation. Written in this form, it is a composition of putting the metric into coordinates corotating with the horizon, and then performing the transformation (2.3). To put the coordinates in null Gaussian form, one would need to also
pass to null coordinates, which regularize the metric at the horizon, before performing the transformation (2.3). Our coordinates are not null, but the metric is diffeomorphic to the one we'd get by performing this extra transformation.\\
We obtain the near-horizon Kerr-AdS(NHEK-AdS) metric:
\begin{equation}
	ds^2 = \frac{\rho_+^2}{V}r^2dt^2 - \frac{\rho_+^2}{V}\frac{dr^2}{r^2} - \frac{\rho_+^2}{f}d\theta^2 - \frac{f\sin^2\theta}{\rho_+^2}\left(\frac{2ar_+}{V}rdt + \frac{r_+^2 + a^2}{\alpha}d\varphi\right)^2
\end{equation}
where we have removed the primes in coordinates, and $\rho_+ = \rho(r_+)$.\\
If we try to perform this transformation on torsion, we see that all the tetrad components of torsion diverge. In fact, one of advantages of using the tetrad formalism, is that this allows us to immediately conclude that the near-horizon
limit of torsion doesn't exist. This is because tetrad components of a tensor are scalar under coordinate transformations. Therefore, if we make a choice of tetrad which is regular in the limit, it is enough to check whether the tetrad components converge.
The tetrad components of torsion are covariant with respect to local Lorentz transformations, and in order to regulate them, a local Lorentz transformation has to be applied, which is necessarily singular at the horizon(this is analogous to the coordinate transformation that we apply when we perform
the continuation of coordinates across the horizon of a black hole). However, this singular transformation will then introduce a divergence in the new choice of tetrads, which won't be regular in the limit. To have a well-defined limit, we are then making a choice of tetrads which are
regular in the limit, along with a partial fixing of Lorentz gauge, allowing only tetrads which are related to our choice by a regular local Lorentz transformation, before taking the limit. This partial fixing of gauge is encoding the geometric conditions for existence of near-horizon limit
of torsion, obtained in section 2. In fact, this procedure can be made general, and unrelated to the particular case of the torsion tensor, and near-horizon limit. For general development of limits of spacetimes, consult \cite{GerochLimits}.\\
Since the divergence that we have discovered cannot be removed, we are unable to use the technique of analyzing near-horizon geometry to obtain extremal black hole entropy for this black hole model. It can be readily checked, that most interesting cases of black holes with dynamical torsion
\cite{BaeklerGuersesHehlMcCreaKerrNewmann1988,CembranosValcarcelRN}, have this same issue. We will devote the rest of the paper to constructing a model where this issue doesn't persist, by looking for a solution with torsion which has the same NHEK-AdS metric, but different form of torsion.

\section{Symmetry ansatz for torsion in NHEK-AdS}
\setcounter{equation}{0}

Since solving the field equations for a new black hole solution with dynamical torsion is quite difficult, we will instead here try to see if there is a solution having the same NHEK-AdS metric, and non-trivial torsion, using the properties of near-horizon geometry that are familiar from GR, in particular
the enhanced symmetry which appears in near-horizon geometries of extremal black holes.\\
Solving the Killing equations for metric (3.11), we obtain the following Killing vectors:
\begin{equation}
\begin{split}
	&\xi_1 = \partial_\varphi \qquad \xi_2 = \partial_t \qquad \xi_3 = t\partial_t - r\partial_r \\
	&\xi_4 = \left(\frac{t^2}{2} + \frac{1}{2r^2}\right)\partial_t - tr\partial_r - \frac{H}{r}\partial_\varphi
\end{split}
\end{equation}
where $H = \frac{2a\alpha r_+}{V(r_+^2 + a^2)}$.\\
The Killing vectors satisfy $SL(2,\mathbb{R})\times U(1)$ symmetry algebra, given by:
\begin{equation}
\begin{split}
    &[\xi_1, \xi_2] = [\xi_1, \xi_3] = [\xi_1, \xi_4] = 0\\
    &[\xi_2, \xi_3] = \xi_2\quad [\xi_2, \xi_4] = \xi_3 \quad [\xi_3, \xi_4] = \xi_4
\end{split}
\end{equation}
The first two of the Killing vectors are inherited from the black hole solution, being related to axisymmetry and stationarity of the solution. The third Killing vector is in fact the generator of the near-horizon transformation. Namely, if we consider the near-horizon transformation (2.3) as one-parameter group
of diffeomorphisms, parametrized by $\varepsilon$, we can easily see that the limit point $\varepsilon \rightarrow 0$ is a fixed point under the compositions of such transformations. Therefore $\xi_3$ as the generator of this transformation automatically becomes Killing vector of the limiting solution. In fact, for any tensor which is
finite in the limit, we find by the same argument that the Lie derivative with respect to $\xi_3$ is equal to zero after the limit, and so this vector generates the symmetry of torsion in near-horizon limit as well. It is a very interesting fact that the near-horizon geometry also generically possesses the
fourth Killing vector which enhances the near-horizon symmetry to $SL(2,\mathbb{R})$ group. This fact, proven in \cite{KunduriLuciettiNHSym,KunduriNH}, isn't guaranteed by the limiting procedure itself, but can be proven using the Einstein equations. Since this symmetry enhancement comes from dynamics, it is model-dependent, and for
generic torsion, we can't expect the $\xi_4$ to be the symmetry of torsion, as it is the case for the metric. In fact, restricting the form of torsion using the first three Killing vectors doesn't change the number of independent components(although it restricts their form), and so field equations with such a form of torsion
remain intractable.\\
For that reason, we are going to assume, in constructing our torsion ansatz, that the torsion possesses the full symmetry of the near-horizon metric, that is, that $\pounds_\xi T^{\mu}_{\hphantom{\mu}\nu\rho} = 0$, for each $\xi \in (\xi_1, \xi_2, \xi_3, \xi_4)$. Solving for independent
components we obtain the following ansatz:
\bsubeq
\begin{align}
	&T^t_{\hphantom{t}t\theta} = T^r_{\hphantom{r}r\theta} =  p_1(\theta) &T^\varphi_{\hphantom{\varphi}t\theta} = -Hr[p_1(\theta) + p_4(\theta)]\\
	&T^t_{\hphantom{t}t\varphi} = T^r_{\hphantom{r}r\varphi} =  p_2(\theta) &T^r_{\hphantom{r}tr} = T^\varphi_{\hphantom{\varphi}t\varphi} = -Hrp_2(\theta) \\
	&T^\theta_{\hphantom{\theta}\theta\varphi} = p_3(\theta) \qquad T^\varphi_{\hphantom{\varphi}\theta\varphi} = p_4(\theta) &T^\theta_{\hphantom{\theta}t\theta}= -Hrp_3(\theta) \\
	&T^\theta_{\hphantom{\theta}tr} = p_5(\theta) \qquad T^\varphi_{\hphantom{\varphi}tr} = p_6(\theta) &T^r_{\hphantom{r}\theta\varphi} = T^\theta_{\hphantom{\theta}t\varphi} = 0
\end{align}
\begin{align}
	&T^t_{\hphantom{t}r\theta} = \frac{1}{r^2}p_7(\theta) \qquad T^t_{\hphantom{t}r\varphi} = \frac{1}{r^2}p_8(\theta) &T^t_{\hphantom{t}tr} = T^\varphi_{\hphantom{\varphi}r\varphi} = -\frac{H}{r}p_8(\theta) \qquad T^\varphi_{\hphantom{\varphi}r\theta} = -\frac{H}{r}p_7(\theta)\\
	&T^r_{\hphantom{r}t\theta} = r^2p_7(\theta) \qquad T^r_{\hphantom{r}t\varphi} = r^2p_8(\theta) &T^t_{\hphantom{t}\theta\varphi} = T^\theta_{\hphantom{\theta}r\theta} = T^\theta_{\hphantom{\theta}r\varphi} = 0
\end{align}
\esubeq
where functions $p_i$, $i = 1 \dots 8$ are arbitrary functions of $\theta$.\\
It is useful to then put the ansatz in tetrad form. We obtain the near-horizon tetrads by taking the near-horizon limit of tetrads given in (3.7). We find:
\begin{equation}
\begin{split}
	&\vartheta^0 = \sqrt{\frac{\rho_+^2}{V}}rdt \qquad \vartheta^1 = \sqrt{\frac{\rho_+^2}{V}}\frac{dr}{r} \\
	&\vartheta^2 = \sqrt{\frac{\rho_+^2}{f}}d\theta \qquad \vartheta^3 = -\sqrt{\frac{f}{\rho_+^2}}\sin\theta\left(\frac{2ar_+}{V}rdt + \frac{r_+^2 + a^2}{\alpha}d\varphi\right)
\end{split}
\end{equation}
The torsion tensor given in this basis takes the following form:
\begin{equation}
\begin{split}
	T^0 &= \Psi_1 \vartheta^0 \wedge \vartheta^2 + \Psi_2 \vartheta^0 \wedge \vartheta^3 + \Psi_3 \vartheta^1 \wedge \vartheta^2 + \Psi_4 \vartheta^1 \wedge \vartheta^3 \\
	T^1 &= \Psi_3 \vartheta^0 \wedge \vartheta^2 + \Psi_4 \vartheta^0 \wedge \vartheta^3 + \Psi_1 \vartheta^1 \wedge \vartheta^2 + \Psi_2 \vartheta^1 \wedge \vartheta^3 \\
	T^2 &= \Psi_5 \vartheta^0 \wedge \vartheta^1 + \Psi_6 \vartheta^2 \wedge \vartheta^3 \\
	T^3 &= \Psi_7 \vartheta^0 \wedge \vartheta^1 + \Psi_8 \vartheta^2 \wedge \vartheta^3
\end{split}
\end{equation}
where now $\Psi_i$, $i = 1\dots 8$, are new independent unknown functions of $\theta$, which are in 1-1 correspondence with the unknown functions from (4.3).\\
The ansatz looks sufficiently simple now, having the same number of undetermined functions as in the case of spherical symmetry \cite{CembranosValcarcelRN,RauchNiehSphTorsion}. Despite this, finding the general solution is not an easy task, and it will be sufficient for now to look for a particular solution where torsion is non-trivial. To do this, we employ
the so called 'double duality' method.

\subsection{Double duality ansatz in PG theory}
Many solutions in PG theory were obtained using the method of double duality, which simplifies the field equations considerably. The method is explained in detail in \cite{ObukhovSelectedTopics,ZhytnikovDoubleDuality}. We will here give a brief recount of the method, before we apply it to our torsion ansatz in the next section.\\
The method is based on the fact that all irreducible components of the curvature tensor satisfy relations of the form:
\begin{equation}
    \prescript{*(n)}{}{R}^{*}_{ij} = \eta_n \prescript{(n)}{}{R}_{ij}
\end{equation}
where $\eta_n = \pm 1$ depending on the particular component. The components are called double-dual for $\eta_n = 1$ and anti-double-dual for $\eta_n = -1$.\\
Here the left star operator is the usual Hodge dual, while the right star operator is the 'right' dual, defined as:
\begin{equation*}
    \prescript{(n)}{}{R}^{*}_{ij} = \frac{1}{2}\varepsilon_{ijkl}R^{kl}
\end{equation*}
A proof of this property can be found in \cite{ObukhovSelectedTopics}.\\
Based on this double duality of curvature components, one can create an ansatz which relates irreducible components of curvature tensor to its double dual. We take here the version of the ansatz presented in \cite{ZhytnikovDoubleDuality}:
\begin{equation}
    \bar{R}_{ij} = \zeta\prescript{*}{}{R}^{*}_{ij} + \chi \vartheta_i \wedge \vartheta_j
\end{equation}
where $\bar{R}_{ij} = \sum_{n=1}^{6}(b_n \prescript{(n)}{}{R_{ij}})$, and $\zeta$ and $\chi$ are undetermined constants.\\
The Hodge dual of the above relation gives a constraint on curvature covariant momenta, and once these relations are substituted into the field equations (3.3), it can be proven after a lengthy calculation that the first equation takes the form of Einstein equation with effective gravitational and cosmologial constants, while the
second equation becomes an algebraic constraint on irreducible components of torsion. In particular, in vaccuum, we find the following set of equations:
\bsubeq
\begin{align}
    -(a_0 -\chi)\varepsilon_{ijkl}\vartheta^j \wedge \tilde{R}^{kl} - 2\Lambda_{\text{eff}}&\varepsilon_{ijkl}\vartheta^j\wedge \vartheta^k \wedge \vartheta^l = 0\\
    (a_1 - a_0 +\chi)\prescript{(1)}{}{T}^i &= 0 \\
    (a_2 + 2a_0 - 2\chi)\prescript{(2)}{}{T}^i &= 0 \\
    (2a_3 + a_0 - \chi)\prescript{(3)}{}{T}^i &= 0
\end{align}
\esubeq
where $\Lambda_{\text{eff}} \equiv \Lambda + \frac{1}{4}\chi R$ is the effective cosmological constant, $(a_0 - \chi)$ is effective gravitational constant, while $\tilde{R}^{ij}$ represents Riemannian curvature, coming from Levi-Civita connection.\\
The double duality relation (4.7) itself, forms a set of algebraic constraints on the irreducible pieces of curvature, written as:
\bsubeq
\begin{align}
    (b_1 + \zeta)\prescript{(1)}{}{R}_{ij} &= 0 \\
    (b_2 - \zeta)\prescript{(2)}{}{R}_{ij} &= 0 \\
    (b_3 + \zeta)\prescript{(3)}{}{R}_{ij} &= 0 \\
    (b_4 - \zeta)\prescript{(4)}{}{R}_{ij} &= 0 \\
    (b_5 + \zeta)\prescript{(5)}{}{R}_{ij} &= 0 \\
    (b_6 + \zeta)R - 12 \chi &= 0
\end{align}
\esubeq
From the relation defining the effective cosmological constant, as well as (4.9f), we see that Ricci scalar has to be a constant, unless $\chi = 0$. When $\chi =0$, all torsion parts are removed from the first field equation and it reduces to the usual Einstein equation. The parameters $a_i$ are related to $a_0$ in such a way as to
reduce the torsion part of the Lagrangian to the one corresponding to the teleparallel equivalent of GR. Thus, in order to obtain a non-trivial solution in PGT, we assume $\chi \neq 0$, and correspodingly, Ricci scalar has to be a constant.\\
The method of finding double duality solutions may be obvious at this point. We pick a metric that satisfies Einstein equations. Then the effective cosmological constant fixes the relation between the Lagrangian parameter of the cosmological constant, and the related metric parameter. The torsion constraints (4.8b)-(4.8d)
allow us to tune the Lagrangian parameters by relating them to the constant $\chi$, leaving them unconstrained for irreducible pieces of torsion which are zero in the solution, and fixing them otherwise. The curvature relations (4.9) provide a similar type of condition on curvature coupling constants in the Lagrangian.
We notice, then, that the freedom in obtaining a double duality solution is considerable, so it is often the case that torsion contains arbitrary functions in such a vaccuum solution. We will find that applying it to our torsion ansatz, we are able to obtain a solution which doesn't have such arbitrariness, therefore
having a potential of containing more interesting physics.

\section{Near horizon Kerr-AdS solution with torsion}
\setcounter{equation}{0}
In this section, we will look for a particular solution to the field equations, assuming that the solution possesses NHEK-AdS metric (3.11), and the torsion is given by the ansatz (4.5). Despite the ansatz having a large symmetry, the Riemann-Cartan curvature components associated with it are already quite complicated. It is safe to
assume that there might be multiple solutions to the problem in different sectors of the theory. Thus, we will try to find a solution that corresponds to the same, or similar sector, as the one in which the solution for Kerr-AdS black hole with torsion is already obtained, given in section 3. This solution has two non-zero irreducible curvature components \cite{BlagojevicCvetkovicKerrAdS}:
\begin{align*}
    \prescript{(4)}{}{R}^{02} &= \prescript{(4)}{}{R}^{12} = \frac{\lambda mr}{\Delta}(b^0 - b^1)\wedge b^2\\
    \prescript{(4)}{}{R}^{03} &= \prescript{(4)}{}{R}^{13} = \frac{\lambda mr}{\Delta}(b^0 - b^1)\wedge b^3\\
    \prescript{(6)}{}{R}^{ij} &= \lambda b^i\wedge b^j
\end{align*}
The only irreducible torsion component that is equal to zero is $\prescript{(3)}{}{T}^i$.
The sector of the theory in which we have this solution is given by the following relations of Lagrangian parameters:
\begin{equation}
    2a_1 + a_2 =0 \qquad a_0 - a_1 - \lambda(b_4 + b_6) = 0 \qquad 3\lambda a_0 + \Lambda = 0
\end{equation}
Out of the six irreducible curvature components, $\prescript{(1)}{}{R}^{ij}$, $\prescript{(4)}{}{R}^{ij}$ and $\prescript{(6)}{}{R}^{ij}$ correspond to the usual components found in GR, being the Weyl tensor, traceless Ricci component, and the Ricci scalar. The other three components $\prescript{(2)}{}{R}^{ij}$, $\prescript{(3)}{}{R}^{ij}$,
$\prescript{(5)}{}{R}^{ij}$ come from the presence of torsion. The first component $\prescript{(1)}{}{R}^{ij}$ is usually the most complicated one, and setting it to zero might be a condition that's too strong, making the system of equations overdetermined. Therefore, in our double-duality approach, we will look for a solution to the system of equations:
\begin{equation}
    \prescript{(2)}{}{R}^{ij} = 0 \qquad \prescript{(3)}{}{R}^{ij} = 0 \qquad \prescript{(5)}{}{R}^{ij} = 0 \qquad \prescript{(3)}{}{T}^{i} = 0
\end{equation}
The torsion equation is solved quite simply, and is given by the following relations:
\begin{equation}
    \Psi_5 = 2\Psi_3 \qquad \Psi_7 = 2\Psi_4
\end{equation}
The second and the third irreducible curvature component are generated by pseudo-Ricci 1-form $X^i$, given in the Appendix A(see also \cite{BlagojevicCvetkovicEntropyHamiltonApproach,ObukhovSelectedTopics}). The fifth irreducible curvature component represents the antisymmetric part of the Ricci tensor. Due to our symmetry assumptions, considering $X^i=0$ gives six independent
constraints, completely restricting the torsion ansatz. It can be checked that this restriction in this case completely corresponds to the first three equations in (5.2) being satisfied. We write the system of equations for $X^i=0$:
\bsubeq
\begin{align}
    \nonumber &F_+\Psi'_2 + (\partial_\theta F_+ + F_+\cot\theta)\Psi_2 + \frac{a r_+ F_+\sin\theta}{\rho_+^2}\Psi_3 +\frac{a^2 F_+\sin\theta\cos\theta}{\rho_+^2}(\Psi_6 - \Psi_2)\\
     &- \Psi_1\Psi_6 - \Psi_2\Psi_8 = 0 \\
    \nonumber &F_+\Psi'_4 + \left(\partial_\theta F_+ + F_+\cot\theta - \frac{a^2F_+\sin\theta\cos\theta}{\rho_+^2}\right)\Psi_4 + \frac{a r_+F_+\sin\theta}{\rho_+^2}(\Psi_1 + \Psi_8)\\
     &-2\Psi_2\Psi_3 + 2\Psi_1\Psi_4 +\Psi_3\Psi_6 + \Psi_4\Psi_8 =0 \\
    &\left(\partial_\theta F_+ + F_+\cot\theta +\frac{a^2F_+\sin\theta\cos\theta}{\rho_+^2}\right)\Psi_4 - \Psi_2\Psi_3 - \Psi_4(\Psi_1 + \Psi_8) = 0 \\
    &F_+\Psi'_3 + \frac{a r_+ F_+\sin\theta}{\rho_+^2}\Psi_6 - \frac{3a^2F_+\sin\theta\cos\theta}{\rho_+^2}\Psi_3 + 2\Psi_1\Psi_3 + \Psi_4\Psi_6 = 0 \\
    &\frac{ar_+F_+\sin\theta}{\rho_+^2}\Psi_2 + \left(\partial_\theta F_+ + F_+\cot\theta\right)\Psi_3 + 2\Psi_2\Psi_4 - \Psi_3\Psi_8 = 0 \\
    &F_+\Psi_4' -\frac{2a^2F_+\sin\theta\cos\theta}{\rho_+^2}\Psi_4 + \frac{ar_+F_+\sin\theta}{\rho_+^2}(\Psi_1 + \Psi_8) + \Psi_2\Psi_3 + \Psi_1\Psi_4 - \Psi_3\Psi_6 = 0
\end{align}
\esubeq
where we have written $F_+ \equiv \sqrt{\frac{f}{\rho_+^2}}$.\\
The derivative of $\Psi_4$ can be eliminated between equations (5.4b) and (5.4f), and we obtain a system of three differential and three algebraic equations. Further, $\Psi_1$, $\Psi_6$ and $\Psi_8$ can be eliminated using the algebraic equations, provided that $\Psi_3 \neq 0$ and $\Psi_4 \neq 0$. We finally get a system of three first order non-linear differential
equations in $\Psi_2$, $\Psi_3$ and $\Psi_4$. It can also be checked that $\Psi_3 = 0$ implies $\Psi_4 = 0$, and $\Psi_4 = 0$ implies $\Psi_3 = 0$. Therefore, the system branches out into two branches: either can $\Psi_3$ and $\Psi_4$ be both zero or both non-zero. Since the non-linear system that appears when both $\Psi_3$ and $\Psi_4$ are non-zero is
quite complicated to solve, we will here look at the degenerate case when $\Psi_3 = \Psi_4 = 0$. The system then simplifies considerably, and we find:
\begin{equation}
    \Psi_1 = - \Psi_8 \qquad \Psi_2 = \Psi_3  = \Psi_4 = \Psi_6 =0
\end{equation}
Torsion tensor now depends only on one function, and we have satisfied the system of equations (5.2). What's left is to impose that the Ricci scalar is constant in this case. We calculate the Ricci scalar to be:
\begin{equation}
    R = 6\left(2\lambda + F_+\Psi_1' + \Psi_1\left(\partial_\theta F_+ + F_+\cot\theta - \frac{2a^2F_+\sin\theta\cos\theta}{\rho_+^2}\right) + \Psi_1^2\right)
\end{equation}
We notice that we can identify the Ricci scalar with that of the known solution by solving the equation:
\begin{equation}
    F_+\Psi_1' + \Psi_1\left(\partial_\theta F_+ + F_+\cot\theta - \frac{2a^2F_+\sin\theta\cos\theta}{\rho_+^2}\right) + \Psi_1^2 = 0
\end{equation}
This equation can be explicitly solved, and we find the closed-form solution to be:
\begin{equation}
    \Psi_1(\theta) = \left(\sqrt{\lambda}a\ln\left(\frac{1+\sqrt{\lambda}a\cos\theta}{1-\sqrt{\lambda}a\cos\theta}\right)-\ln\left(\frac{1+\cos\theta}{1-\cos\theta}\right) + c\right)^{-1}\frac{2\alpha}{\rho_+\sqrt{f}\sin\theta}
\end{equation}
where $c$ is the integration constant.\\
We have thus obtained a solution with non-trivial torsion that has the metric of NHEK-AdS solution. We first calculate its irreducible components of curvature and torsion, in order find the sector of the theory in which the solution resides. The non-zero irreducible components of curvature are given by:
\bsubeq
\begin{align}
    &\prescript{(1)}{}{R}^{01} = -2C \vartheta^0 \wedge \vartheta^1 - 2D \vartheta^2 \wedge \vartheta^3\\
    &\prescript{(1)}{}{R}^{02} = C \vartheta^0 \wedge \vartheta^2 - D \vartheta^1 \wedge \vartheta^3 \qquad \prescript{(1)}{}{R}^{03} = C \vartheta^0 \wedge \vartheta^3 + D \vartheta^1 \wedge \vartheta^2\\
    &\prescript{(1)}{}{R}^{12} = C \vartheta^1 \wedge \vartheta^2 - D \vartheta^0 \wedge \vartheta^3 \qquad \prescript{(1)}{}{R}^{13} = C \vartheta^1 \wedge \vartheta^3 + D \vartheta^0 \wedge \vartheta^2\\
    &\prescript{(1)}{}{R}^{23} = -2C \vartheta^2 \wedge \vartheta^3 + 2D \vartheta^0 \wedge \vartheta^1
\end{align}
\esubeq
where:
\begin{equation*}
    C = \frac{mr_+}{\rho_+^6}(r_+^2 - 3a^2\cos^2\theta) \qquad D = \frac{ma\cos\theta}{\rho_+^6}(3r_+^2 - a^2\cos^2\theta)
\end{equation*}
\bsubeq
\begin{align}
    &\prescript{(4)}{}{R}^{01} = P\vartheta^0 \wedge \vartheta^1\\
    &\prescript{(4)}{}{R}^{02} = -Q \vartheta^0 \wedge \vartheta^2 \qquad \prescript{(4)}{}{R}^{03} = Q \vartheta^0 \wedge \vartheta^3\\
    &\prescript{(4)}{}{R}^{12} = -Q \vartheta^1 \wedge \vartheta^2 \qquad \prescript{(4)}{}{R}^{13} = Q \vartheta^1 \wedge \vartheta^3\\
    &\prescript{(4)}{}{R}^{23} = -P \vartheta^2 \wedge \vartheta^3
\end{align}
\esubeq
where:
\begin{align*}
    P &= \Psi_1(\theta)\left(\Psi_1(\theta) - \frac{2a^2 F_+ \sin\theta\cos\theta}{\rho_+^2}\right)\\
    Q &= \Psi_1(\theta)\left(\Psi_1(\theta) + \partial_\theta F_+ + F_+\cot\theta - \frac{a^2F_+\sin\theta\cos\theta}{\rho_+^2}\right)
\end{align*}
\begin{equation}
    \prescript{(6)}{}{R}^{ij} = \lambda \vartheta^i \wedge \vartheta^j
\end{equation}
The only irreducible component of torsion that is non-zero is $\prescript{(2)}{}{T}^i$, so we have $T^i = \prescript{(2)}{}{T}^i$.\\
We see that the sector of the solution is different from the one presented in section 3. We calculate the relations between Lagrangian parameters to be:
\begin{equation}
    b_1 = -b_4 \qquad a_2 + 2a_0 - 2\lambda(b_4 + b_6) = 0 \qquad \Lambda + 3\lambda a_0 =0
\end{equation}
The important question that now remains is whether this solution corresponds to a near-horizon limit of a black hole with torsion. We answer this question in the next section.

\section{Towards the black hole solution}
\setcounter{equation}{0}

The solution we have obtained for torsion in the last section is given by:
\bsubeq
\begin{align}
    T^0 &= \Psi(\theta)\vartheta^0 \wedge \vartheta^2 \\
    T^1 &= \Psi(\theta)\vartheta^1 \wedge \vartheta^2 \\
    T^2 &= 0\\
    T^3 &= -\Psi(\theta)\vartheta^2 \wedge \vartheta^3
\end{align}
\esubeq
where $\Psi(\theta)$ is given by equation (5.8).\\
We are looking for a black hole that corresponds to this solution, that is, a Kerr-AdS black hole with torsion, with metric and tetrad defined by (3.4) and (3.7), and torsion which reduces to (6.1) after taking the near-horizon limit in the extremal case. This black hole now obviously doesn't have to possess the
complete symmetry found in near-horizon solutions, but instead we would require it only to be stationary and axisymmetric. Thus, we can expect that there are in principle multiple possible black holes which have the same near-horizon limit.\\
In order to minimize the ambiguities in choosing the torsion generalization, we will attempt to generalize our solution in the following form:
\bsubeq
\begin{align}
    T^0 &= \Psi(r, \theta)b^0 \wedge b^2 \\
    T^1 &= \Psi(r, \theta)b^1 \wedge b^2 \\
    T^2 &= 0\\
    T^3 &= -\Psi(r, \theta)b^2 \wedge b^3
\end{align}
\esubeq
where tetrads are now coming from the black hole metric, given by (3.7).\\
The change of tetrads and form of function $\Psi$ in the ansatz will induce changes in curvature components, and we may expect new terms to appear. The main thing that we wish to check first is the change of Ricci scalar, which we have to maintain to be constant in order to apply the double duality method.
We calculate the Ricci scalar to be:
\begin{equation}
    R = 6\left(2\lambda + F(\partial_\theta\Psi) + \Psi\left(\partial_\theta F + F\cot\theta - 2\frac{a^2 F\sin\theta\cos\theta}{\rho^2}\right) + \Psi^2\right)
\end{equation}
where $F = \sqrt{\frac{f}{\rho^2}}$ and $\rho = \sqrt{r^2 + a^2\cos\theta}$, which are functions appearing in the black hole metric.\\
We obtained the same form of the equation as we did in the near-horizon solution. Solving for $R = 12\lambda$ as before, we obtain:
\begin{equation}
    \Psi(r,\theta) = \left(\sqrt{\lambda}a\ln\left(\frac{1+\sqrt{\lambda}a\cos\theta}{1-\sqrt{\lambda}a\cos\theta}\right)-\ln\left(\frac{1+\cos\theta}{1-\cos\theta}\right) + c(r)\right)^{-1}\frac{2\alpha}{\rho\sqrt{f}\sin\theta}
\end{equation}
We see that what appeared before as the integration constant is now an arbitrary function of radial coordinate $c(r)$. We may anticipate that this arbitrary function will be unphysical(as is usually the case), and that actual solution of interest will have $c(r) = 0$(and likewise $c \equiv c(r_+) = 0$ in the
near-horizon limit), however we leave it here for completeness. Its meaning will be revealed in future research, when the complete analysis of the solution will be performed.\\
Since we managed to set Ricci scalar to constant, and that torsion in fact correctly corresponds to the one found in section 5 after taking the near-horizon limit, we proceed to calculate other irreducible torsion and curvature components. We may expect that the change in ansatz may induce the appearance of
new non-zero irreducible components, and in fact this will be the case.\\
The irreducible components of torsion are again simple, and we have:
\begin{equation}
    \prescript{(1)}{}{T}^i = 0 \qquad \prescript{(2)}{}{T}^i = T^i \qquad \prescript{(3)}{}{T}^i = 0
\end{equation}
The non-zero irreducible curvature components read:
\bsubeq
\begin{align}
    &\prescript{(1)}{}{R}^{01} = -2C \vartheta^0 \wedge \vartheta^1 -\Psi\frac{rN}{2\rho^2} \vartheta^0 \wedge \vartheta^2 - 2D \vartheta^2 \wedge \vartheta^3\\
    &\prescript{(1)}{}{R}^{02} = -\Psi\frac{rN}{2\rho^2} \vartheta^0 \wedge \vartheta^1 + C \vartheta^0 \wedge \vartheta^2 - D \vartheta^1 \wedge \vartheta^3 \qquad \prescript{(1)}{}{R}^{03} = C \vartheta^0 \wedge \vartheta^3 + D \vartheta^1 \wedge \vartheta^2\\
    &\prescript{(1)}{}{R}^{12} = C \vartheta^1 \wedge \vartheta^2 - D \vartheta^0 \wedge \vartheta^3 \qquad \prescript{(1)}{}{R}^{13} = C \vartheta^1 \wedge \vartheta^3 + D \vartheta^0 \wedge \vartheta^2 + \Psi\frac{rN}{2\rho^2} \vartheta^2 \wedge \vartheta^3\\
    &\prescript{(1)}{}{R}^{23} = -2C \vartheta^2 \wedge \vartheta^3 + 2D \vartheta^0 \wedge \vartheta^1 + \Psi\frac{rN}{2\rho^2} \vartheta^1 \wedge \vartheta^3
\end{align}
\esubeq
where:
\begin{equation*}
    C = \frac{mr}{\rho^6}(r^2 - 3a^2\cos^2\theta) \qquad D = \frac{ma\cos\theta}{\rho^6}(3r^2 - a^2\cos^2\theta)
\end{equation*}
\bsubeq
\begin{align}
    &\prescript{(2)}{}{R}^{01} = -\Psi\frac{rN}{2\rho^2} \vartheta^0 \wedge \vartheta^2 \qquad \prescript{(2)}{}{R}^{02} = \Psi\frac{rN}{2\rho^2} \vartheta^0 \wedge \vartheta^1 \qquad \prescript{(2)}{}{R}^{03} = 0\\
    &\prescript{(2)}{}{R}^{12} = 0 \qquad \prescript{(2)}{}{R}^{13} = \Psi\frac{rN}{2\rho^2} \vartheta^2 \wedge \vartheta^3 \qquad \prescript{(2)}{}{R}^{23} = -\Psi\frac{rN}{2\rho^2} \vartheta^1 \wedge \vartheta^3
\end{align}
\esubeq
\bsubeq
\begin{align}
    &\prescript{(4)}{}{R}^{01} = P\vartheta^0 \wedge \vartheta^1 + \frac{1}{2}N(\partial_r\Psi)\vartheta^0 \wedge \vartheta^2 - \Psi\frac{aN\cos\theta}{\rho^2}\vartheta^1 \wedge \vartheta^3\\
    &\prescript{(4)}{}{R}^{02} = \frac{1}{2}N(\partial_r\Psi)\vartheta^0 \wedge \vartheta^1 -Q \vartheta^0 \wedge \vartheta^2 - \Psi\frac{aN\cos\theta}{\rho^2}\vartheta^2 \wedge \vartheta^3\\
    &\prescript{(4)}{}{R}^{03} = Q \vartheta^0 \wedge \vartheta^3 \qquad \prescript{(4)}{}{R}^{12} = -Q \vartheta^1 \wedge \vartheta^2\\
    &\prescript{(4)}{}{R}^{13} = \Psi\frac{aN\cos\theta}{\rho^2}\vartheta^0 \wedge \vartheta^1 + Q \vartheta^1 \wedge \vartheta^3 + \frac{1}{2}N(\partial_r\Psi)\vartheta^2 \wedge \vartheta^3\\
    &\prescript{(4)}{}{R}^{23} = \Psi\frac{aN\cos\theta}{\rho^2}\vartheta^0 \wedge \vartheta^2 + \frac{1}{2}N(\partial_r\Psi)\vartheta^1 \wedge \vartheta^3 -P \vartheta^2 \wedge \vartheta^3
\end{align}
\esubeq
where:
\begin{align*}
    P &= \Psi(r, \theta)\left(\Psi(r, \theta) - \frac{2a^2 F \sin\theta\cos\theta}{\rho^2}\right)\\
    Q &= \Psi(r, \theta)\left(\Psi(r, \theta) + \partial_\theta F + F\cot\theta - \frac{a^2F\sin\theta\cos\theta}{\rho^2}\right)
\end{align*}
\bsubeq
\begin{align}
    &\prescript{(5)}{}{R}^{01} = -\frac{1}{2}N(\partial_r\Psi) \vartheta^0 \wedge \vartheta^2 \qquad \prescript{(5)}{}{R}^{02} = \frac{1}{2}N(\partial_r\Psi) \vartheta^0 \wedge \vartheta^1 \qquad \prescript{(1)}{}{R}^{03} = 0\\
    &\prescript{(5)}{}{R}^{12} = 0 \qquad \prescript{(1)}{}{R}^{13} = -\frac{1}{2}N(\partial_r\Psi) \vartheta^2 \wedge \vartheta^3 \qquad \prescript{(5)}{}{R}^{23} = \frac{1}{2}N(\partial_r\Psi) \vartheta^1 \wedge \vartheta^3
\end{align}
\esubeq
\begin{equation}
    \prescript{(6)}{}{R}^{ij} = \lambda \vartheta^i \wedge \vartheta^j
\end{equation}
We notice that new irreducible components have indeed appeared, and that the limit of the irreducible components correctly corresponds to the near-horizon solution that we have already obtained. The appearance of new components means that the sector of solution is going to be more restricted, compared to the near-horizon
limit, which is to be expected. The sector of solutions can only shrink or remain the same when the limit is 'lifted', corresponding to the fact that multiple black holes may in principle have the same near-horizon limit. We calculate the sector of solution to be:
\begin{equation}
    b_1 = -b_2 = -b_4 = b_5 \qquad a_2 + 2a_0 - 2\lambda(b_4 + b_6) = 0 \qquad \Lambda + 3\lambda a_0 =0
\end{equation}
We have now obtained a Kerr-AdS black hole with non-trivial torsion which has a regular near-horizon limit. The analysis of  this solution is left for future studies.

\section{Discussion}

We have examined the general problem of identifying near-horizon geometries of extremal black holes with torsion. It appears that most black hole solutions, which are the most complex in Poincaré gauge theory, do not possess a regular near-horizon limit in the extremal case. By finding a solution that exhibits this property, we have opened the door for further investigation. Many questions still remain unanswered. 

We developed a general method for deriving near-horizon limits of extremal black holes with torsion. While we interpreted the geometric conditions we imposed, a deeper analysis could yield additional insights. Notably, three out of four conditions impose restrictions on the torsion current at the horizon. These conditions could be further interpreted in the context of horizon stationarity with non-trivial torsion, similar to how the first condition was discussed in \cite{ZerothLawTorsion}. The fourth condition is differential in nature and resembles the extremality condition imposed on the metric in general relativity (GR), which ensures the existence of a near-horizon limit. A more thorough understanding of this condition may enhance our comprehension of extremality in geometries with torsion. 

The near-horizon geometry model we developed was based on extending known near-horizon symmetries to include torsion tensor symmetries. This approach is expected to produce results consistent with the double duality method, as the field equations simplify to Einstein-like equations—an important aspect for proving such symmetries in GR \cite{KunduriLuciettiNHSym,KunduriNH}. However, the full system of equations was not solved in the most general case. Further exploration might reveal solutions that are less distinct from those obtained by Baekler et al. \cite{McCreaBaeklerGuersesPGKerr,BaeklerGuersesHehlMcCreaKerrNewmann1988}. Additionally, the black hole solution presented in this work warrants a comprehensive analysis within the canonical approach. Since the sector in which the black hole was constructed is more restricted than that of the near-horizon solution, calculations of entropy in the extremal case need to be confined to this sector. Ideally, this could be addressed by modifying the torsion ansatz for the black hole, adding terms that vanish in the near-horizon limit, thereby aligning the black hole solution sector with the near-horizon sector. However, this is a delicate process, as introducing extra torsion terms may significantly alter the curvature tensor.

We hope that addressing these questions in future research will deepen our understanding of black hole geometries with torsion. This could lead to improved physical interpretations of these geometries and their relation to the well-established results from standard GR.

\section*{Acknowledgments}

This work was  supported by the Ministry of  Science, Technological Development and Innovation  of the Republic of Serbia.

\appendix

\section{Irreducible decomposition of curvature and torsion tensors}
\setcounter{equation}{0}
We present here the formulas for irreducible components of curvature and torsion in 4D spacetime. These formulas are the same as the ones given in \cite{BlagojevicCvetkovicEntropyHamiltonApproach}. We have labeled the orthonormal frame dual to the tetrad 1-form $\vartheta^i$, by $e_i$, the defining relation being $e_i {\inn} \vartheta^j = \delta^j_i$.\\
The torsion 2-form has three irreducible components:
\begin{align}
    &\nonumber\prescript{(2)}{}{T}^i = \frac{1}{3}\vartheta^i \wedge (e_m {\inn} T^m)\\
    &\nonumber\prescript{(3)}{}{T}^i = \frac{1}{3}e^i {\inn} (T^m \wedge \vartheta_m)\\
    &\prescript{(1)}{}{T}^i = T^i - \prescript{(2)}{}{T}^i - \prescript{(3)}{}{T}^i
\end{align}
The curvature 2-form decomposes into six irreducible parts:
\begin{align}
    &\nonumber\prescript{(2)}{}{R}^{ij} = \prescript{*}{}{(\vartheta^{[i} \wedge W^{j]})} &&\prescript{(4)}{}{R}^{ij} = \vartheta^{[i} \wedge V^{j]}\\
    &\prescript{(3)}{}{R}^{ij} = \frac{1}{12}X\prescript{*}{}{(\vartheta^i \wedge \vartheta^j)} &&\prescript{(6)}{}{R}^{ij} = \frac{1}{12}R \vartheta^i \wedge \vartheta^j\\
    &\nonumber\prescript{(5)}{}{R}^{ij} = \frac{1}{2}\vartheta^{[i} \wedge e^{j]} {\inn} (\vartheta^m \wedge Ric_i) &&\prescript{(1)}{}{R}^{ij} = R^{ij} - \sum_{m=2}^{6}\prescript{(m)}{}{R}^{ij} 
\end{align}
where:
\begin{align}
    &\nonumber Ric^i \equiv e_j {\inn} R^{ji} \qquad R \equiv e_i {\inn} Ric^i \\
    &X^i \equiv \prescript{*}{}{(R^{ik} \wedge \vartheta_k)} \qquad X \equiv e_i {\inn} X^i
\end{align}
and:
\begin{align}
    &\nonumber V^i = Ric^i - \frac{1}{4}R\vartheta^i - \frac{1}{2}e^i {\inn} (\vartheta^m \wedge Ric_m)\\
    &W^i = X^i - \frac{1}{4}X\vartheta^i - \frac{1}{2}e^i {\inn} (\vartheta^m \wedge X_m)
\end{align}

\end{document}